\documentstyle[preprint,epsf,aps]{revtex} 

\def\hea{He~{\scriptsize I}}
\def\heb{He~{\scriptsize II}}
\begin{document}
%\twocolumn[\hsize\textwidth\columnwidth\hsize\csname@twocolumnfalse\endcsname
\draft
\preprint{Draft - not for distribution \ \ \ \ \today}

\title{\bf Electronic structure of NiS$_{1-x}$Se$_x$} 

\author{S.~R.~Krishnakumar, N.~Shanthi, Priya~Mahadevan\cite{jrcat} and
D.~D.~Sarma\cite{jnc}} 

\address{Solid State and Structural  Chemistry Unit, 
Indian Institute of Science, Bangalore, 560012, India}
\maketitle 
\begin{abstract}
%\vspace{0.2in} 

We investigate the electronic structure of the metallic NiS$_{1-x}$Se$_x$ system
using various electron spectroscopic techniques. The band structure results do
not describe the details of the spectral features in the experimental spectrum,
even for this paramagnetic metallic phase.  However, a parameterized many-body
multi-band model is found to be
successful in describing the Ni~2$p$ core level and valence band, within the
same model.
The asymmetric line shape as well as the weak intensity feature in
the Ni~2$p$ core level spectrum has been ascribed to extrinsic loss processes in
the system.  The presence of satellite features in the valence band spectrum
shows the existence of the lower Hubbard band, deep inside the $pd$ metallic
regime, consistent with the predictions of the dynamical mean field theory.

\end{abstract}

\vspace{0.2in} 


%]
%\narrowtext 

\section{Introduction}

Electronic structure of 3$d$-transition metal compounds has been an active area
of research for many decades, owing to their rich and diverse physical
properties, often exhibiting exotic transport and magnetic properties.  In
particular, the electronic structure of the hexagonal nickel sulphide~(NiS) has
attracted a lot of interest \cite{nis_prl,Nakamura1} due to the fact that it has
a first-order phase transition as a function of temperature at about 260~K.  The
high-temperature phase is a Pauli paramagnetic metal, transforming
isostructurally to an antiferromagnetic state with nearly temperature
independent resistivity \cite{Koehler1,Coey1}.  The high temperature phase is
metallic with a resistivity in the order of 10$^{-5}$~Ohm-cm 
\cite{Koehler1,Coey1} which is comparable with other highly metallic inorganic
compounds of 3$d$ transition elements \cite{springer}.  The metallic state can
be stabilized by the application of pressure or substitution with Se for S.  A
pressure of more than 20~kbar \cite{Mcwhan1} or a Se percentage of more than
13\% \cite{Anzai1} suppresses the electronic phase transition, stabilizing the
metallic phase down to the lowest temperature. At room temperature, the hexagonal 
NiS crystallizes in the NiAs structure having
space group $P6_3/mmc$ ($D^4_{6h}$) and lattice parameters, $a$ = 3.440~\AA\ and
$c$ = 5.351~\AA\ \cite{Trahan}.  
The Ni$^{2+}$ is in an octahedral environment of
sulphur atoms in NiS and these NiS$_6$ octahedra are edge-shared within the
$ab$-plane and face-shared along the $c$-axis, forming a three-dimensional
network of Ni-S connectivity in this system.

In spite of its interesting properties, there have been few studies to
understand the electronic structure of these compounds.  Band structure
calculations \cite{Mattheiss1,Fujimori4,Raybaud} indeed yield a metallic ground
state for the high-temperature phase. 
However, the details of the electronic structure, as shown
in the present work, cannot be entirely described within the band theory due to the
presence of substantial correlation effects even in the metallic phase.
Different configuration interaction models have been employed in order to
explain the valence band \cite{Fujimori1} and the transition metal core level
spectra \cite{Bocquet}; however, there has not been any attempt to describe both
using a single model and with a single set of parameters.  In the present paper,
we investigate the electronic structure of metallic NiS$_{1-x}$Se$_x$ system using
$x$-ray photoemission (XP), ultra-violet photoemission (UP), and bremsstrahlung
isochromat (BI) spectroscopic measurements in conjunction with {\it ab-initio}
band structure calculations as well as parameterized many-body calculations, in
order to provide a consistent and quantitative description of the electronic
structure and the electronic parameter strengths controlling the physical
properties of this system.

\section{EXPERIMENTAL}

Polycrystalline samples of NiS$_{1-x}$Se$_x$ with $x$ = 0.0 and 0.15 used for
the present study were prepared following the standard solid state reaction
techniques reported in the literature \cite{nis_prl,Anzai1}; 
$x$-ray diffraction patterns as well as resistivities of the samples 
were found to be in agreement with the
reported data in the literature \cite{Anzai1,Trahan}.
Spectroscopic measurements were carried out in a combined VSW spectrometer with
a base pressure of 2$\times10^{-10}$ mbar equipped with a monochromatized
Al~K$\alpha$ $x$-ray source, He discharge lamp and an electron gun.  XP, BI, and
UP spectroscopic measurements were performed on the samples with an overall
instrumental resolution of better than 0.8~eV, 0.8~eV, and 90~meV, respectively.
BI spectra have been recorded by monochromatizing the emitted photons at
1486.6~eV.  The sample surface was cleaned {\it in situ} periodically during the
experiments by scraping with an alumina file and the surface cleanliness was
monitored by recording the carbon 1$s$ and oxygen 1$s$ XP signals.  The
reproducibility of the spectral features was confirmed in each case.  The
binding energy was calibrated to the instrumental Fermi-level which was
determined by recording the Fermi-edge region from a clean silver sample. 
Experimental spectra of NiS reported here
were collected for the highly conducting state at room temperature ($\sim
295$~K), while those of NiS$_{0.85}$Se$_{0.15}$, which does not exhibit any
transition, were collected at $\sim 120$~K.  Using high resolution photoemission
spectroscopy, it has been established \cite{nis_prl} that the electronic
structure close to $E_F$ changes across the electronic transition at $\sim
260$~K.  However, our low-resolution spectroscopic results show hardly any
change in the spectra across the transition either in the valence band or in the
core levels, in agreement with previous studies \cite{Fujimori1,Matoba}
establishing similar gross electronic structures with comparable electronic
parameter strengths for both phases across the electronic transition.

Scalar relativistic linearized muffin-tin orbital (LMTO) band structure
calculations have been performed within the atomic sphere approximation (ASA)
for the paramagnetic phase of NiS with the real crystal structure.  The unit
cell has two formula units and the sphere radii for Ni and S were 2.4 and
3.12~au, respectively.  In this calculation, convergence was obtained
self-consistently using $s$, $p$ and $d$ basis at each atomic sphere with
186~$k$ points in the irreducible part of the Brillouin Zone.  In order to
understand and compare the experimental electronic structure with that obtained
from the band structure calculations, it is necessary to include the effects of
photoemission cross-sections of various levels.  Thus, the theoretical spectra
based on the band structure calculations were constructed by multiplying the
various partial densities of states (DOS) with the corresponding cross-sections
computed within the formalism of ref.~\cite{winter}.  The cross-section weighted
DOS was then convoluted with a Lorentzian with energy dependent full-width at
half maximum, to account for the lifetime broadening and a Gaussian to describe
the instrumental resolution broadening.

Core level and valence band~(VB) spectra were calculated for a NiS$_6$ cluster
with the geometry in accordance with the bulk crystal structure of this
compound, within a parameterized many-body multi-band model including orbital
dependent electron-electron (multiplet) interactions \cite{PMT,Dimen}.  The
multiplet interactions within the $3d$ manifold were expressed in terms of 
Slater integrals $F^0_{dd}$, $F^2_{dd}$ and $F^4_{dd}$ and those within the
$2p-3d$ manifold were expressed in terms of $F^0_{pd}$, $F^2_{pd}$, $G^1_{pd}$ and
$G^3_{pd}$.  The values $F^2_{dd}$ (=9.79~eV), $F^4_{dd}$ (=6.08~eV), 
$F^2_{pd}$ (=6.68~eV), $G^1_{pd}$ (=5.07~eV) and
$G^3_{pd}$ (=2.88~eV) were taken to be 80\% of the Ni atomic Hartree-Fock values. 
The monopole terms, $F^0_{dd}$ and $F^0_{pd}$ were
treated as parameters and therefore, varied to fit the experimental spectrum.
It is often more convenient to discuss the electron-electron interaction
strengths in terms of the multiplet averaged quantities, $U_{dd}$ and $U_{pd}$,
where

\begin{equation}
U_{dd} = F^0_{dd} - \frac{2}{63}(F^2_{dd} + F^4_{dd}), 
\end{equation}
\begin{equation}
U_{pd} = F^0_{pd} - \frac{1}{15}G^1_{pd} - \frac{3}{70}G^3_{pd}. 
\end{equation}

\noindent The multiplet averaged $U_{pd}$ term was set to have a value 1.2 times
that of $U_{dd}$ \cite{Dimen}, thus reducing the number of freely adjustable
parameters in the model.  The hopping interaction strengths between Ni~3$d$ and
S~3$p$ were expressed in terms of the Slater-Koster parameters $(pd\sigma)$ and
$(pd\pi)$ and that between the different bonding S~3$p$ orbitals were expressed
in terms of $(pp\sigma)$ and $(pp\pi)$ \cite{Slater}.  We fixed the ratio
between $(pd\pi)$ and $(pd\sigma)$ to be -0.5, while that between $(pp\pi)$ and
$(pp\sigma)$ to be -0.2, in these calculations.

The calculations were performed in the symmetry adapted ($t_{2g}$ and $e_g$)
basis including the transition metal 3$d$ and the bonding sulphur 3$p$ orbitals.
In the calculation for the valence band spectrum, the S~3$p$ spectral
contribution was also calculated within the same parameterization.  As the size
of the Hamiltonian is large, the Lanczos method was used to evaluate the
spectral function and the calculated one-electron removal spectra were
appropriately broadened to simulate the experimental spectra.  In the Ni~2$p$
core level calculation, Doniach-Sunjic line shape function was used for
broadening the discrete energy spectrum of the cluster model, in order to
represent the asymmetric line shape of core levels from these highly metallic
compounds, consistent with other core levels in the system.  In the case of VB
calculations, energy dependent Lorentzian function was used for the lifetime
broadening.  Other broadening effects such as the resolution broadening and
solid state effects were taken into account by convoluting the spectra with a
Gaussian function.  The broadening parameters were found to be consistent with
values used for similar systems \cite{Dimen}.  Taking into account the different
atomic cross-sections for the Ni~3$d$ and S~3$p$ states, it is necessary to
calculate a weighted average of these two contributions to the valence band.
The atomic cross-section ratio \cite{Yeh} between S $3p$ and Ni $3d$ states 
($\approx$ 0.17) is not appropriate
in this context, since solid-state effects alter this ratio significantly. 
It was found that S 3$p$/Ni 3$d$ cross-section ratio of approximately
5.5 times that obtained from the atomic calculations gives the best result
for the cluster calculations.

\section{RESULTS \& DISCUSSIONS}

The total density of states (DOS) for NiS along with the Ni~$d$ and S~$p$
partial DOS obtained from the band structure calculations for the paramagnetic
phase are shown in Fig.~1.  The calculated DOS is in good agreement with
previous reports \cite{Fujimori4,Raybaud}.  The Fermi level lies close to a
minimum in the DOS, and there is a finite DOS at the $E_F$, consistent with the
metallic nature of the system.  The total DOS between -3 and 1~eV is dominated
by Ni~$d$ contributions with the energy region around the most intense peak in
the DOS at about -1.65~eV dominated by the Ni~$t_{2g}$ states and the features
at higher energies ($\ge -1$~eV) arising primarily from the $e_g$ states.  As a
consequence of the high covalency of the system, the Ni~3$d$ ($e_g$) states have
substantial contribution from S~3$p$ states.  These states constitute the
antibonding states due to Ni~$d$-S~$p$ interactions.  Their bonding counterparts
are located between -5 and -3~eV and have dominant S~3$p$ contributions.  The
non-bonding states of the system, centered around -7~eV is weaker in intensity
and is stabilized at a higher binding energy compared to Ni~$d$-S~$p$ bonding
states, which is a generic feature as seen in the case of the transition metal
monosulphides \cite{Raybaud}.

We compare the experimental XP valence band and BI spectra corresponding to the
occupied and unoccupied parts of the electronic structures for NiS and
NiS$_{0.85}$Se$_{0.15}$ in Fig.~2(a) and (b), respectively.  The calculated
spectra for NiS, obtained using various partial DOS 
as explained in the previous section, 
are also shown for comparison. 
The XP and BI spectra of NiS and NiS$_{0.85}$Se$_{0.15}$ show finite
spectral intensity at $E_F$, indicating their highly metallic conductivity.  It
is clear from the experimental spectra in Figs.~2(a) and (b) that the
substitution of Se in place of S does not cause any perceptible change in the XP
valence band spectra as well as in the BI spectra.  This indicates that the
occupied as well as the unoccupied parts of the electronic structure in these
two compounds are similar, in agreement with the similar magnetic and transport
properties of the two compounds.

Fig.~2(a) shows that the peak in the experimental spectra around 1.3~eV binding
energy (marked A) essentially consists of Ni~$3d$ states, while that at
$\sim4$~eV binding energy (marked B) is due to the S~3$p$ contribution.
However, the calculated spectrum does not describe the experimental spectra
accurately, as the energy positions of the different features, A and B are not
correctly predicted by the calculation, but are shifted to higher binding
energies compared to the experiment. 
Additionally, the experimental feature A is narrower compared to the calculated one. 
Moreover, the structure around 7~eV
binding energy (marked C) in the experimental spectra is absent in the
theoretical spectrum.  In the unoccupied part of the electronic structure, the
peak at around -0.7~eV in the experimental BI spectra is contributed mainly by
the unoccupied e$_g$ states of the Ni~$3d$ manifold, strongly hybridized with
the S~$3p$ bands, as seen from Fig.~1.  The strong suppression in the S~3$p$
contribution in the calculated spectrum (Fig.~2(b)) compared to the partial DOS
(Fig.~1) in the vicinity of $E_F$ arises from a substantially lower
photoemission cross-section for the S~3$p$ states compared to Ni~3$d$ at Al
K$\alpha$ $x$-ray energies.  The broad spectral features above 4~eV are due to
the various high energy unoccupied states in the system, such as Ni~$4s, 4p$ and
S~$3d$.  Similar to the case in XP spectra, the band structure calculation,
while providing a reasonable description of the overall spectral features
observed in the BI spectra, fail to reproduce the energy position of the feature
A in Fig.~2(b) by approximately 0.3~eV.  These failures of the band theory to
explain the electronic structure of this system indicates that electron
correlation effects are important even in the paramagnetic metallic phase.

The valence band spectra of NiS and NiS$_{0.85}$Se$_{0.15}$ recorded using
different photon energies, namely 1486.6~eV (XPS), 40.8~eV (\heb) and 21.2~eV
(\hea) are compared in Fig.~3.  The various features are marked as A, B and C
and represent the same features as in Fig.~2(a).  At all the photon energies,
there is a finite spectral intensity at $E_F$.  
Besides
providing a much higher resolution compared to the XPS, UP spectra with \hea\
and {\scriptsize II} radiations allow for a wider variation in the relative
photoemission cross-sections of Ni~3$d$ and S~3$p$, leading to an easy
identification of their relative contributions to the valence band spectra.  
The marked loss in
the intensity of the feature  B  for a moderate increase of the photon energy
between \hea\ and \heb\ radiations arises from the Cooper minimum in the S~$3p$
photoionization cross-section at around 50~eV, 
clearly establishing the dominance of S~3$p$ contributions in the
energy range of the feature B, with feature A having predominantly Ni~3$d$
contributions, in agreement with previous works \cite{Fujimori1}.  
The valence band spectra for NiS and NiS$_{0.85}$Se$_{0.15}$
taken at various photon energies are very similar, showing the similarity in
their electronic structure within the solid solution, NiS$_{1-x}$Se$_x$ for
$x\le0.15$.

The transition metal $2p$ core level spectra in transition metal compounds often
reveal important information on their electronic structures.  The Ni~$2p$ core
level spectra for NiS and NiS$_{0.85}$Se$_{0.15}$ are shown in the  inset of
Fig.~4; the two spectra are very similar to each other.  Each spectrum consists
of spin-orbit split, $2p_{3/2}$ and $2p_{1/2}$ peaks at 853.4~eV and 870.7~eV
binding energies, respectively with pronounced satellite features at 860~eV and
876~eV, suggesting that electron correlations are important in the system.  The
satellite intensity relative to the main peak is considerably more in the
$2p_{1/2}$ region compared to that in the $2p_{3/2}$ region.  While it is indeed
possible to have different shapes and intensity for the two satellite features
accompanying $2p_{3/2}$ and $2p_{1/2}$ features due to details of the multiplet
interactions, such a strong variation between the two satellites as observed
here is unexpected.  In order to explore the possibility of strong variations in
the inelastic background function, normally assumed to vary smoothly, we have
performed electron energy loss spectroscopy (EELS) on these samples, with the
same primary energy as that of the Ni~$2p$ core level peak.  The inelastic loss
background appropriate for the Ni~2$p$ photoemission spectrum was constructed
from the experimental EELS spectrum by convoluting it with a Gaussian to obtain
similar width for the elastic peak as that in the core level main peak.  The
region near the elastic peak was replaced by an integral background function.
The total inelastic background was then generated by adding the loss functions
corresponding to the $2p_{3/2}$ and $2p_{1/2}$ features with the intensities
given by the degeneracy ratio (2:1) of the core levels.  The resulting
background function is shown in the inset of Fig.~4 as a solid line.  We find
that there is an intense and structured contribution to the background function
arising from a plasmon band overlapping the $2p_{1/2}$ satellite region and
resulting in the anomalously large satellite intensity in the $2p_{1/2}$ region
compared to that in the $2p_{3/2}$ region.  It is also seen that at around
856~eV, there is a peak-like structure in the inelastic background; this appears
at the same position as that of a weak intensity feature at about the same
energy in the core level experimental spectrum as shown by the vertical dashed 
lines in the inset of Fig.~4.  This structure in the inelastic
scattering background could have its origin from the inter-band $p-d$
transitions.  Thus, it appears that this weak intensity shoulder in the
experimental Ni~2$p$ spectrum is largely due to this extrinsic loss feature in
the background function. 
Our explanation is contrary to the earlier interpretation
\cite{Bocquet} of this feature as arising from intrinsic processes. 

In order to obtain a quantitative description of the electronic structure of
this system, we have calculated the Ni~$2p$ core level and valence band (VB)
spectra within a single model involving a NiS$_6$ cluster, retaining the local
geometry of the real solid.  
The calculated Ni~$2p$ spectrum (solid line) for the
parameter set S-Ni $(pd\sigma)$ = -1.4~eV, S-S $(pp\sigma)$ = 0.7~eV, $\Delta$ =
2.5~eV and $U_{dd}$ = 4.0~eV, including the experimentally determined inelastic
background is superposed on the experimental spectrum (open circles) in the main
panel of Fig.~4, exhibiting a good agreement; the calculated spectrum without
any broadening is presented as bar diagrams.  There is an underestimation of the
intensity around the shoulder to the main peak near 857~eV and around 883~eV
binding energies.  These small discrepancies are possibly due to the differences
between the experimental EELS spectrum used for generating the background
function and the actual inelastic loss background for the photoemission
spectrum.  The previous estimates \cite{Bocquet} of the various parameter
strengths in NiS obtained from a model that included a fictitious conduction band  in
addition to Ni~3$d$-S~3$p$ basis within the cluster model used for the core
level calculation, are $(pd\sigma)$ = -1.47~eV, $\Delta$ = 2.5~eV, and $U_{dd}$
= 5.5~eV.  Thus, the present estimates differ significantly only for $U_{dd}$,
for which we have obtained a smaller value.

The $(pd\sigma)$ values estimated for NiS is similar to that estimated for other
divalent nickel sulphides, for example, $(pd\sigma)$ = -1.5~eV for NiS$_2$ and
BaNiS$_2$ \cite{Unpublished}, and for trivalent Ni oxide systems, such as
LaNiO$_3$ ($(pd\sigma)$ = -1.57~eV \cite{sudipta}).  However, divalent Ni oxide systems, 
such as NiO, typically exhibit a Ni~3$d$-O~2$p$ $(pd\sigma)$ = -1.25~eV \cite{Dimen}.
This decrease in the strength  of $(pd\sigma)$ for the divalent
oxides compared to trivalent ones is due to an increase in the Ni-O distances in
the divalent oxides.  The similarity in the $(pd\sigma)$ of nickel sulphides and
the trivalent Ni oxides, in spite of the differences in their Ni-ligand bond
length ($d_{Ni-O}\simeq$~1.93~\AA\ for LaNiO$_3$ compared to
$d_{Ni-S}\simeq$~2.39~\AA\ in NiS)  arises from the fact that S~3$p$
orbitals are much more spatially extended in nature than O~2$p$ and hence have a
larger overlap integral with Ni~3$d$ orbitals. 
In general,
$\Delta$ is expected to be smaller for sulphides compared to oxides, since the
O~$2p$ levels are more stable than the S~$3p$ levels; for example, the estimated
$\Delta$ for NiO is 5.5~eV \cite{Dimen} compared to 2.5~eV in NiS. 
The value of $U_{dd}$ for NiS (4.0~eV) is much smaller than that
in NiO (6.5~eV) possibly due to a more efficient screening in NiS, arising from
the smaller charge transfer energy, $\Delta$ and larger hopping strength,
$(pd\sigma)$, leading to stronger covalency effects.

We have analyzed the ground state wavefunction of NiS corresponding to the
 estimated parameter strengths in terms of the various electron configurations.
 The ground state of the system was found to consist of 60.7\%, 35.8\% and 3.4\%
 of $d^8$, $d^9\underline{L}^1$, and $d^{10}\underline{L}^2$ configurations with
 a high-spin configuration ($S = 1$).  The average value of the $d-$occupancy
 $(n_d)$ is found to be 8.43, showing a highly covalent ground state of the
 system.  In the case of NiO, a typical divalent oxide of Ni, $n_d$ is found to
 be $\sim8.16$ \cite{Dimen}.  We have analyzed the characters of the final
 states of the system responsible for different features in the experimental
 spectrum, in order to understand their origins.  The analysis was carried out
 for the final state energies marked 1-10 in Fig.~4.  The different
 contributions to the final states from various electron configurations ($d^8$,
 $d^9\underline{L}^1$, $d^{10}\underline{L}^2$) are listed in Table.~I.  These
 features can be grouped into three different regions; the main peak region,
 852-856~eV (labeled 1-3 in Fig.~4); intense satellite region, 858-861~eV
 (labeled 4-6), and weak satellites in the region 863-866~eV (labeled 7-10).
 The first group of features in the main peak region has a dominant
 $d^9\underline{L}^1$ character as seen from the table, which are the
 ``well-screened" states of the system and correspond to one ligand (sulphur)
 electron being transferred to the Ni site to screen the Ni~$2p$ core-hole
 potential created by the photoemission process. 
The second group of features, has a
 mixed character with significant contributions from all the configurations.
 This is in contrast to the case of NiO where the intense satellite structure
 results primarily from the ``poorly-screened" $d^8$ configuration.  
The third group of features
 has a dominant $d^{10}\underline{L}^2$ contribution 
for the lower energy region (labeled 7 and 8) and a
 dominant $d^8$ character for the higher energy region (labeled 9 and 10 in
 Fig.~4), with a considerable amount of $d^9\underline{L}^1$ character for the
 entire region.

The XP valence band spectrum of NiS has also been calculated within the same
model.  An important feature of the present calculation is that the S~$3p$
contribution to the VB spectrum has been calculated on the same footing as the
contribution from the Ni~3$d$ within the {\it same} model without the need to
adjust either the shape or the energy position of the ligand contribution to the
valence band region, thereby enhancing the reliability of the estimated
parameter strengths.  The calculated spectrum (solid line) along with the
Ni~$3d$ (dashed line) and S~$3p$ (dot-dash line) contributions for NiS are shown
superimposed on the experimental data (open circles) in Fig.~5.  An inelastic
scattering background function (dotted line) is also included in the total
calculated spectrum.  The calculated contributions from the Ni~$3d$ to the total
spectrum for NiS is also shown without any broadening effect as an energy stick
diagram in Fig.~5.  The parameter set used for the VB calculation is similar to
that used for the core level calculation, but with a Ni~3$d$-S~3$p$ $(pd\sigma)$
of -1.2~eV instead of -1.4~eV.  Such minor adjustments of parameter strengths
between the simulations for the core and valence level photoemission spectra,
are of common occurrence.  The present estimates of the electronic structure
parameters are similar to that obtained from the previous valence band
calculations \cite{Fujimori1} within the cluster-model. 
The agreement
between the experimental spectrum and the calculated spectrum in the present
study (see Fig.~5) is remarkable over the entire energy range.  The main peak in
the valence band spectrum in Fig.~5 at about 1.3~eV arises essentially from
Ni~3$d$ photoemission contribution, though there is a small contribution arising
from S~3$p$ states also due to the covalent  mixing of states.  It is possible 
to delineate such contributions only in the present
approach where S~3$p$ contribution is also calculated within the same model.
Within the traditional approach, S~3$p$ contribution would be approximated by a
single Gaussian with its peak position, width and intensity being freely
adjusted to obtain the best possible description.  Clearly the present approach
is more reliable and satisfactory.  A second, and more intense, S~3$p$
contribution is responsible for the experimental spectral feature near 4~eV
binding energy.  The shoulder appearing between about 6 and 9~eV is once again
dominated by the Ni~3$d$ contributions.  This observation along with the fact
that this feature is not in agreement with the {\it ab-initio} band structure
results (see Fig.~2), clearly suggests that this feature is driven by
correlation effects.  The analysis of the ground state wavefunction for the
parameter set used for the VB spectrum shows that the ground state has 64.9\%,
32.6\% and 2.5\% of $d^8$, $d^9\underline{L}^1$ and $d^{10}\underline{L}^2$
characters giving rise to an average $d-$occupancy ($n_d$) of 8.38 electrons
with a high-spin configuration for the Ni$^{2+}$, similar to that obtained from
the core level calculation. 
The parameter strengths estimated here place NiS$_{1-x}$Se$_x$ samples 
in $pd$-metal region of the ZSA phase diagram \cite{ZSA,Seva}.

 The results of the character analysis of the final
states labeled 1-11 in the figure corresponding to the Ni~$3d$ contribution in
the VB spectrum of NiS are shown in Table.~II.  On the basis of this analysis,
the spectral features can be grouped into three regions, the main peak region
(0-6~eV, labeled 1-6), the strong satellites in the 6-8.5~eV range (labeled 7-9)
and weaker satellites beyond 8.5~eV (marked 10 and 11).  The final states in the
main peak region predominantly consist of $d^8\underline{L}^1$ states with a
non-negligible contributions from $d^7$ and $d^9\underline{L}^2$ configurations.
This is similar to the scenario as observered in the case of other
charge-transfer systems, like NiO.  In the satellite region (6-8.5~eV), the
final states are dominated by $d^7$ and $d^9\underline{L}^2$ configurations.
The weak intensity features at higher energies are dominated by
$d^9\underline{L}^2$ configuration with a substantial contribution also from
$d^8\underline{L}^1$ states. 
It is to be noted that the first ionization state, marked 1 in the figure, 
is not at the Fermi energy. This is 
the well known artifact of finite cluster-model 
calculations which invariably exhibit 
a gap in the charge excitation spectrum 
due to the discrete eigen-spectrum of a 
finite cluster. It is believed that this 
ionization state acquires a band width in the 
thermodynamic limit closing the 
discrete gap and overlaping the $E_F$. 
A dispersion width of less than 1~eV is seen to
be sufficient to bring this about; such a dispersion is consistent with our 
results obtained from LMTO band structure calculations for the infinite solid. 
While we have not performed an analysis of the orbital symmetry, our analysis 
for the first ionization state shows it to be a spin-doublet state.

It is evident that the shoulder-like structure at around 7~eV in the
experimental VB spectra is primarily due to the satellite structure of Ni~$3d$
states and is expected to have some non-bonding S~$3p$ contribution as seen from
its photon energy dependence (see Fig.~3).  This explains the inability of the
band structure calculations to explain this feature, as the origin of this
feature is primarily due to the electron correlations present in the system,
giving rise to strong satellite structures.  
The same correlation effects are also responsible for shifting the main peak to 
a lower energy and making it narrower compared to the band structure results 
(see Fig.~3).  This satellite structure, with a
dominant $d^7$ character, is the spectral signature of the lower Hubbard band.
This observation along with the highly conducting Pauli paramagnetic property of
these compounds suggests that the Hubbard side-bands exist deep inside the
metallic regime.  This is in accordance with the predictions of the dynamical
mean field theory (DMFT) calculations based on the Hubbard model \cite{Georges},
that well defined Hubbard bands continue to exist in the spectral functions even
far away from the Mott transition in the metallic region, though the
Mott-Hubbard gap collapses by the transfer of spectral weight from the Hubbard
side-bands to the coherent spectral features near $E_F$ region.

In conclusion, we have investigated the electronic structure of the system,
NiS$_{1-x}$Se$_x$, using various electron spectroscopic techniques to elucidate
the underlying electronic structure.  The substitution of Se in place of S does
not show appreciable changes in the electronic structure of the system,
indicating similar electronic parameters for this solid solution.  The band
structure results do not describe well the details of the spectral features in
the paramagnetic phase of the system, while a parameterized many-body multi-band
model 
is found to be successful in describing
the core level and valence band spectrum of the NiS.  The core and valence band
spectral calculations of paramagnetic metallic phase of NiS show that it is a
correlated metal with a highly covalent character.  The asymmetric line shape as
well as the weak feature in the core level spectra have been ascribed to
extrinsic loss processes in the system.  The valence band calculations show that
lower Hubbard band exists well inside the $pd$ metallic regime, as predicted by
the DMFT calculations.  The electronic parameter strengths obtained for the
compounds of this solid solution show that they belong to the $pd$ metallic
regime of the ZSA phase diagram.

\section{Acknowledgments}

The authors thank Professor C.  N.  R.  Rao for continued support and the
Department of Science and Technology, and the Board of Research in Nuclear Sciences, 
Government of India, for financial
support.  SRK  thanks the Council of Scientific and Industrial Research,
Government of India, for financial assistance. DDS thanks Dr. M. Methfessel, 
Dr. A. T. Paxton, and Dr. M. van Schiljgaarde for making the LMTO-ASA band 
structure program available. The authors also thank Professor
S.  Ramasesha and the Supercomputer Education and Research Center, Indian
Institute of Science, for providing the computational facility.

\pagebreak

\section{figure captions}

Fig.~1.  Density of states (DOS) of NiS from the band structure calculation for
the paramagnetic state.  The thick solid line represents the total DOS.  The
thin solid and dashed lines are the Ni~3$d$ and S~3$p$ partial DOS,
respectively.

Fig.~2.  (a) Experimental valence band spectra using Al K$\alpha$ for NiS and
NiS$_{0.85}$Se$_{0.15}$ along with the 
calculated spectrum obtained from band
structure calculations.  The thick solid line represents the total spectrum,
while the thin solid and dashed lines correspond to the contributions to the
calculated spectrum from Ni~3$d$ and S~3$p$ partial DOS, respectively. 
(b) Experimental BI
spectra for NiS and NiS$_{0.85}$Se$_{0.15}$ along with calculated spectrum
obtained from the band structure calculations.  Various contributions to the
total spectrum are represented by different line types.

Fig.~3.  Experimental valence band spectra for NiS and NiS$_{0.85}$Se$_{0.15}$
using 1486.6~eV (XPS), 40.81~eV (\heb) and 21.2~eV (\hea) photon energies.
Various features in the spectra are shown by the vertical lines and are labeled
as A, B, and C (see text).

Fig.~4.  Experimental Ni~$2p$ spectrum (open circles) along with the calculated
spectrum (solid line) for NiS obtained from the cluster calculation.  Various
final states of the cluster calculation and the corresponding intensity
contributions without any broadening are shown as the bar diagram.  Inset shows
Ni~$2p$ core level spectra for NiS and NiS$_{0.85}$Se$_{0.15}$ along with the
inelastic scattering background function obtained from EELS for NiS.  The dashed
vertical line shows the structure in the background function responsible for the
feature in the core level spectra around 856~eV.

Fig.~5.  The experimental VB spectrum (open circles) along with the calculated
spectrum (solid line), Ni~$3d$ component (dashed line), S~$3p$ component
(dot-dashed line) and the integral background (doted line) are shown for NiS.
The final states of the calculation and the corresponding intensities without
any broadening are shown as the energy stick diagram.

\onecolumn 

\begin{table}

TABLE I.~~~Contributions from various configurations in the final states of the
Ni 2$p$ core level photoemission in NiS.  The peak numberings correspond to the
labels indicated in Fig.~4; the corresponding binding energies (B.E) in eV are
also shown.  \\

\begin{tabular}{c|c c c c c c c c c c}
Peak no. & 1 & 2 & 3 & 4 & 5 & 6 & 7 & 8 & 9 & 10 \\ B.E & 853.4
&  854.1  & 854.7  & 859.1  & 859.7 & 860.9 & 863.3 & 863.7 & 865.0 &
865.9 \\ \hline 

$d^8$ & 23.62 & 12.83 & 10.77 & 42.47 & 
25.62 & 32.90 & 17.43 & 16.80 & 52.44 & 63.44  \\ 

$d^9\underline{L}^1$ & 57.51
& 61.87 & 61.02 & 24.39 &  50.86 & 17.87  & 37.93 & 35.80 &
34.74 & 28.42  \\ 

$d^{10}\underline{L}^2$  & 18.87 &  25.30 &
28.21 & 33.14 &  23.52 & 49.23 & 44.65 & 47.40 & 12.82 & 8.14 \\ 
\end{tabular}
\end{table}

\begin{table}

TABLE II.~~~Contributions from various configurations in 
the final states of the
valence-band photoemission in NiS.  The peak numberings correspond to the labels
indicated in Fig.~5; the corresponding binding energies (B.E) in eV are also
shown.  \\

\begin{tabular}{c|c c c c c c c c c c c}
Peak no. & 1 & 2 & 3 & 4 & 5 & 6 & 7 & 8 & 9 &  10 & 11\\ 

B.E & 0.8 & 1.3&  2.4 & 3.6 & 4.5  & 5.5 & 6.5 &  7.0 & 7.7 &
9.0 & 11.1\\ \hline 

$d^7$ & 12.23 & 26.12 & 15.23 & 0.00  & 0.43 & 1.39  & 51.91 & 30.73 &
44.40 & 12.22 & 9.71\\ 

$d^8\underline{L}^1$ & 57.07 & 57.13 & 60.15 & 78.46 & 72.52 & 77.24 &
10.46 & 19.32 & 7.64 & 27.31 & 36.88 \\ 

$d^9\underline{L}^2$ & 28.35 & 16.16 & 23.56 & 21.54 & 25.98 & 20.79 &
34.78 & 40.57 & 43.69 & 60.19 & 50.48 \\ 

$d^{10}\underline{L}^3 $ & 2.35 & 0.59 & 1.06  & 0.00 & 1.07 &
0.58 & 2.85 & 9.38 & 4.27 & 0.28 & 2.93 \\ 
\end{tabular}
\end{table}

\end{document}